\documentclass[prd,amsmath,amssymb,twocolumn,superscriptaddress,reprint,floatfix,aps]{revtex4-2}
\usepackage{graphicx}
\usepackage{bm}
\usepackage[utf8]{inputenc}
\usepackage[T1]{fontenc}
\usepackage{mathptmx}
\usepackage{etoolbox}
\usepackage{verbatim}
\usepackage{xcolor}
\usepackage{ulem}
\usepackage{orcidlink}
\usepackage{comment}

\usepackage{hyperref}
\hypersetup{colorlinks,
 linkcolor=blue,%
 citecolor=blue,%
 urlcolor=blue
}

\begin{document}
\title{Anomalous Energy Injection in the Gross-Pitaevskii Framework for Turbulence in Neutron Star Glitches}

\author{Anirudh Sivakumar\orcidlink{0009-0007-9527-4555}}%
\affiliation{Department of Physics, Bharathidasan University, Tiruchirappalli 620 024, Tamil Nadu, India}

\author{Pankaj Kumar Mishra\orcidlink{0000-0003-1566-0012}}
\affiliation{Department of Physics, Indian Institute of Technology Guwahati, Guwahati 781039, Assam, India}

\author{Ahmad A. Hujeirat\orcidlink{0000-0001-5178-6240}}
\affiliation{Interdisciplinary Center for Scientific Computing, The University of Heidelberg, 69120 Heidelberg, Germany}

\author{Paulsamy Muruganandam\orcidlink{0000-0002-3246-3566}}
\affiliation{Department of Physics, Bharathidasan University, Tiruchirappalli 620 024, Tamil Nadu, India}
\affiliation{Department of Medical Physics, Bharathidasan University, Tiruchirappalli 620 024, Tamil Nadu, India}

\begin{abstract}
Neutron star glitches---sudden increases in rotational frequency---are thought to result from angular momentum transfer via quantized vortices in the superfluid core. To investigate the underlying superfluid dynamics, we employ a two-dimensional rotating atomic Bose-Einstein condensate described by a damped Gross-Pitaevskii equation with an imposed pinning potential that serves as a simplified analogue of a crust. Within this minimal framework, we examine the emergence and evolution of turbulent vortex motion following impulsive perturbations reminiscent of glitch-like forcing. Our simulations reveal a transient Kolmogorov-like turbulent cascade ($k^{-5/3}$) that transitions to a Vinen-like scaling ($k^{-1}$). We identify an anomalous secondary injection mechanism driven primarily by quantum pressure, which can sustain turbulent fluctuations in such a system. By tuning the damping coefficient $\gamma$, we determine an optimal regime for energy transfer. While idealized, these findings illustrate how quantum turbulence with multiple scaling regimes can arise in pinned, rotating superfluids, and they suggest possible qualitative connections to vortex-mediated dynamics in neutron stars and other astrophysical superfluid systems.
\end{abstract}

\date{\today}
\maketitle

\textit{Introduction:} Decaying neutron stars exhibit sudden increases in rotation frequency, known as glitches~\cite{Radhakrishnan1969}, caused by the transfer of angular momentum from quantized vortices in the neutron superfluid to the outer crust of the pulsar~\cite{Manchester2017, Hujeirat2018}. This transfer is primarily triggered by the vortex avalanche mechanism, in which quantum vortices become pinned to crustal nuclei~\cite{Chamel2008}, preventing the superfluid from spinning down at the same rate as the crust. The resulting rotational lag between the crust and superfluid generates a Magnus force on the vortices~\cite{Sonin1997}. Once a critical velocity is reached, the quantized vortices unpin from their pinning sites~\cite{Stockdale2021,Schwarz1981}. These vortices are then expelled toward the crust, transferring their angular momentum and producing a glitch~\cite{Melatos2008}. This glitch process involves a complex vortex avalanche encompassing millions of vortices, triggering their depinning~\cite{Loennborn2019}. 

Modeling the depinning avalanche mechanism poses significant challenges. Vortex depinning initiates when a vortex, initially trapped by a pinning site, overcomes the attractive force and begins moving in the direction of the applied superfluid flow. Several factors, such as the critical velocity of the superfluid flow and the size and shape of the pinning site, affect the overall dynamics of depinning~\cite{Liu2024}. However, an alternative analysis suggests that glitches emerge due to a quantum transition in which the core transforms into a superconducting gluon-quark superfluid state, with resultant vortex depinning occurring at the boundary layer between the core and the surrounding dissipative medium~\cite{Hujeirat2018}. In this Letter, we demonstrate anomalous energy transfer from quantum pressure to incompressible kinetic energy, thereby proposing a candidate mechanism for turbulent dynamics within a Gross-Pitaevskii (GP) analog of neutron star glitches.

Rotating Bose-Einstein condensates (BECs) provide a quantitatively precise, experimentally tunable, and microscopically resolvable analogue for investigating vortex avalanche processes in neutron-star superfluids \cite{Migdal1959, Warszawski2011, Warszawski2012}. Verma {\it et al.}~\cite{Verma} and Shukla {\it et al.} \cite{Shukla} developed a minimal model for the emergence of the superfluid glitches, based on the interaction of the neutron-superfluid vortices and proton-superconductor flux tubes. More recently, Poli \textit{et al.}~\cite{Poli2023} extended Gross–Pitaevskii models of pinned superfluids by including dipolar interactions and realistic crustal pinning potentials, demonstrating the emergence of self-organised criticality in simulated glitch events. Although these and related Gross–Pitaevskii–based approaches have successfully reproduced many observed features of pulsar glitches, they remain limited by the modest number of vortices and the small physical size of the simulated condensates; limitations that are unavoidable with current computational resources and that do not affect more phenomenological (non-microscopic) vortex-avalanche models. Building on this Gross–Pitaevskii framework, our study investigates the turbulent superfluid dynamics triggered by large vortex avalanches, uncovering a previously overlooked anomalous energy-injection mechanism.

The inherent compressibility of BECs renders them susceptible to quantum turbulence (QT) under dynamic instabilities. This turbulence is primarily characterized by the breakdown of vortices in a self-similar process, leading to a power-law scaling of $k^{-5/3}$, known as the Kolmogorov spectrum \cite{Kobayashi2007, Kobayashi2008}. In addition to the Kolmogorov regime, QT in BECs also manifests in the Vinen regime, which is characterized by random vortex distributions in superfluids and exhibits a $k^{-1}$ scaling. Recent developments have identified novel turbulence regimes, including strong quantum turbulence \cite{Barenghi2023, MiddletonSpencer2023}, marked by density fluctuations and non-polarized vortex lines, as well as rotational quantum turbulence \cite{Estrada2022, Estrada2022a, Sivakumar2024, Sivakumar2024a}, and turbulence in self-gravitating dark-matter BEC candidates \cite{Mocz2017, Sivakumar2025}.

Previous studies have explored the nature of glitches and vortex avalanches in the context of pinning potentials and superfluid behavior. However, the role of superfluid-crust interactions in inducing turbulent flow remains largely unexamined. In this Letter, we investigate the characteristics of turbulence and associated velocity flows arising from such interactions. Furthermore, it analyzes the mechanisms driving this turbulence, facilitated by pinning sites and the dynamic spin-down of the condensate. The damping effects on the system are also evaluated, identifying an optimal damping coefficient that maximizes turbulence strength. Recent studies indicate that quantum turbulence suppresses collective excitation modes in BECs \cite{Lee2025, Ferrand2021}. As these modes drive vortex avalanches and glitch phenomena, quantifying the onset of turbulence in such systems is essential. Beyond its implication for existing neutron star models, the effect of vortex depinning, particularly its role in driving incompressible turbulence through secondary injection, is of significant interest in the domain of quantum turbulence~\cite{Bradley2006, Henn2009, Neely2013, Reeves2013, Barenghi2023}.

\textit{The Model:}  
We model the interior of a spinning-down neutron star's superfluid core using a quasi-two-dimensional (quasi-2D) atomic Bose-Einstein condensate. This rotating BEC system, incorporating dissipative interactions, is described by the damped Gross-Pitaevskii equation, presented in its non-dimensional form in Ref.~\cite{Liu2025}.
\begin{align}\label{eq:gpe}
(\mathrm{i} - \gamma) \frac{\partial \psi}{\partial t} = \left[ -\frac{1}{2} \nabla^2 + V(\mathbf{r}, t) + g |\psi|^2 - \Omega(t) L_z \right] \psi,
\end{align}
where $\psi \equiv \psi(\mathbf{r}, t)$ is the condensate wave function, with $\mathbf{r} \equiv (x, y)$. 
$\nabla^2 = \partial_x^2 \, + \partial_y^2$ represents the two-dimensional Laplacian, $\gamma$ corresponds to the parameterized damping coefficient, and $g$ is the nonlinear interaction strength given by $g = 800$. 
{Although the interaction parameter does not directly represent the microscopic nuclear interactions in a neutron superfluid, it models a short-range repulsive potential that captures the essential physics for generating quantum vortices. For the large-scale hydrodynamic behavior and collective vortex dynamics relevant to glitch-scale turbulence, the system can be described by effective theories where microscopic details are parameterized (e.g., into a mutual friction coefficient), rather than acting as the primary driver of the dynamics \cite{Chamel2017}.}

The time-dependent rotation frequency $\Omega(t)$ denotes the rotational deceleration profile of the condensate which have the forms as $\Omega(t) = \Omega_0 \cos^2 \left( \frac{\pi t}{2 t_s} \right)$ for $t \leq t_s$ and zero for  $t > t_s$. Here $\Omega_0$ is the initial rotation frequency of the condensate, and $t_s$ is the spin-down time. The $z$-component of the angular momentum is given by $L_z = \mathrm{i} \hbar (y \partial_x - x \partial_y)$. 
In general, glitch events arise from transitions between two distinct non-zero rotational frequencies~\cite{Andersson2012, Poli2023}. 
However, a partial spin-down can retain vortices (see Fig. \ref{fig:author:density} in Appendix \ref{sec:appendix_a}), which may suppress the amplitude of glitches~\cite{supp}. While introducing a sudden deceleration can overcome this suppression, it often results in an unstable and unconfined condensate. Therefore, to enhance the vortex avalanche process without destabilizing the condensate, we have instead chosen to fully spin down the condensate.

The potential term $V(\mathbf{r},t)$ comprises the following components
\begin{align}
V(\mathbf{r},t) = V_{\mathrm{box}}(\mathbf{r}) + V_{\mathrm{crust}}(\mathbf{r},t) + V_{\mathrm{cent}}(\mathbf{r},t).
\end{align}
The confining circular-box potential, $V_{\mathrm{box}}$, with radius $R_{\mathrm{box}}$, is defined as $V_{\mathrm{box}}(\mathbf{r}) = V_{0b}\Theta(r - R_{\mathrm{box}})$~\cite{Liu2025}, where $\Theta$ is the Heaviside step function and $r = \sqrt{x^2 + y^2}$. For our analysis, we set the barrier height and radius of the circular-box trap as $V_{0b} = 100$ and $R_{\mathrm{box}} = 6$, respectively. The crust potential, which acts as a vortex pinning site, is given by~\cite{Shukla, Verma}:
\begin{align}
V_{\mathrm{crust}}(\mathbf{r},t) = V_{0c} \exp\left( -\frac{(|\mathbf{r}_p| - r_{\mathrm{crust}})^2}{(\Delta r_{\mathrm{crust}})^2} \right) \tilde{V}(x_\theta, y_\theta),
\label{eq:crust}
\end{align}
where $\tilde{V}(x_\theta, y_\theta) = 3 + \cos(n_{\mathrm{crust}} x_\theta) + \cos(n_{\mathrm{crust}} y_\theta)$, and the rotated coordinates are $x_\theta = \cos(\theta(t)) x_p + \sin(\theta(t)) y_p$ and $y_\theta = -\sin(\theta(t)) x_p + \cos(\theta(t)) y_p$. Here, $n_{\mathrm{crust}}$ determines the number of pinning sites, $r_{\mathrm{crust}}$ is the radius at which $V_{\mathrm{crust}}$ is maximized, and $\Delta r_{\mathrm{crust}}$ represents the crust thickness. The crust potential parameters are based on Refs.~\cite{Shukla, Verma}. The centrifugal potential, $V_{\mathrm{cent}}(\mathbf{r},t) = \frac{1}{2} \Omega(t)^2 r^2$, ensures uniform condensate density under rotation. Following Refs.~\cite{Liu2025, Graber2017, Seveso2016}, we adopt a coherent length scale (comparable to vortex core size) of $\xi = 10~\mathrm{fm}$, yielding an energy scale of $\epsilon = 207~\mathrm{keV}$. With these parameters, the time scale is calculated as $\tau = 3.2 \times 10^{-21}~\mathrm{s}$. We compute the kinetic energy components and their spectra based on the analysis of their numerical implementation in Ref.~\cite{Bradley2022}. 
It is important to note that while the GP model is inherently limited to describing weakly interacting BECs in relatively small systems, resulting in a mismatch with the microscopic length scales present in neutron stars, it effectively captures vortex depinning phenomena, which are considered strong candidates for characterizing glitch events. Although scaling the vortex population, condensate radius and rotation frequency to astrophysical scales is unfeasible, the GP model has reproduced Self-Organized Criticality of the pulsar glitch behavior ~\cite{Poli2023,Shukla}
The simulation involves numerically solving the GP equation \eqref{eq:gpe} using the split-step Crank-Nicolson method \cite{Muruganandam2009, Kumar2019} in a computational domain of size $512 \times 512$, with spatial steps $dx = dy = 0.05$ and a time step $dt = 10^{-3}$ to ensure numerical stability and convergence. The numerical simulations are performed on an NVIDIA A100 GPU using CUDA C codes developed based on~\cite{Loncar2016}.
Initially, we prepare the condensate and crust with an initial rotation frequency $\Omega_0 = 2$ through imaginary-time iterations ($t \to -\mathrm{i} t$). This converged profile is then spun down at varying rates, characterized by the spin-down time $t_s$, in real-time. We first consider a case with a static crust and no dissipation ($\gamma = 0$) while exploring the effects of damping in subsequent analyses. 

\begin{figure}[!ht]
\centering
\includegraphics[width=\linewidth]{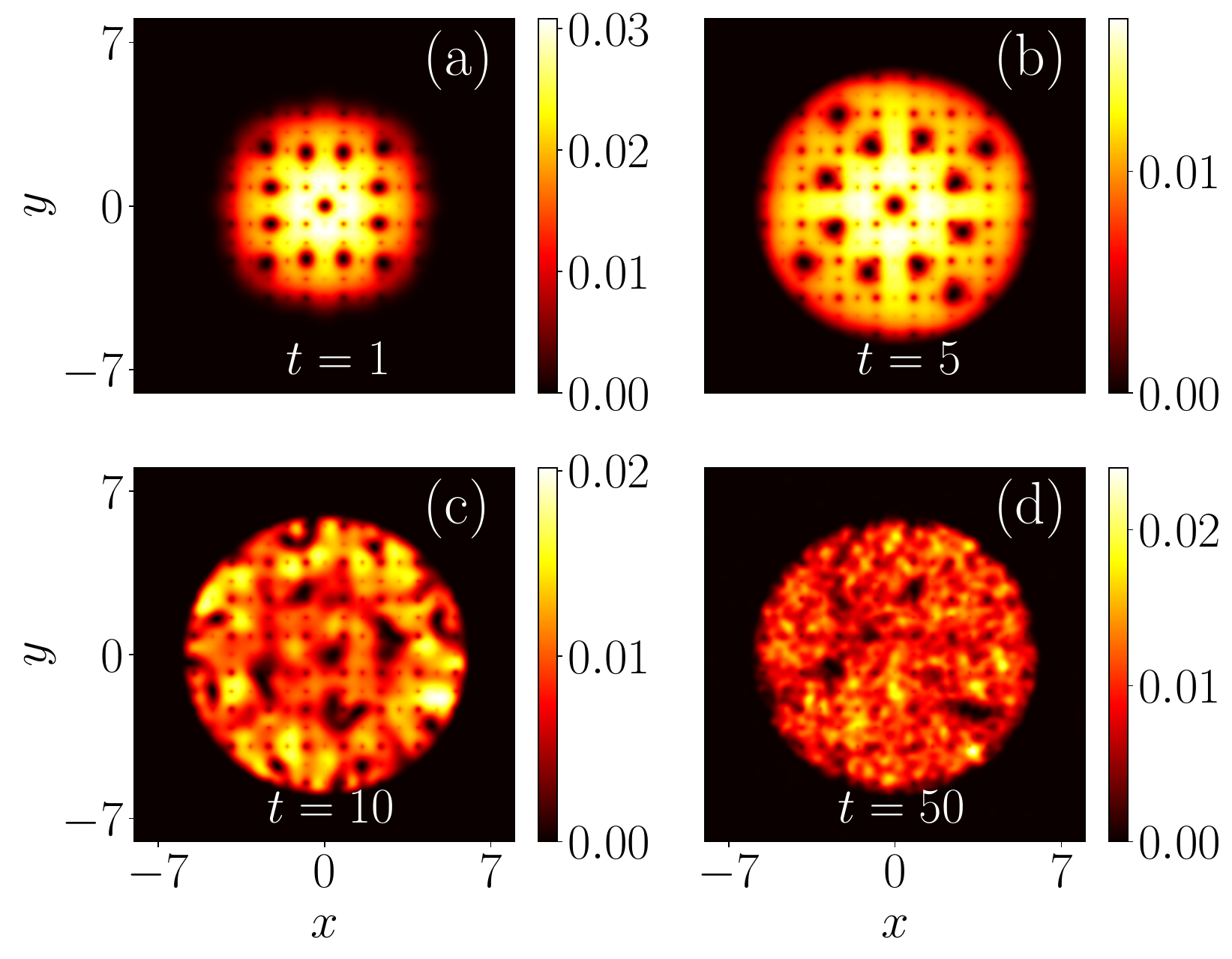}
\caption{Snapshots of condensate density during real-time spin-down at $t_s = 10$ with the crust potential. As the condensate spins down, the initially dominant centrifugal confinement [(a)] is overtaken by the circular box trap [(b)--(d)]. Concurrently, turbulent flow induces the depinning of vortices.}
\label{fig:denprofile_nr}
\end{figure}

\textit{Results:} 
Superfluidity significantly impacts the interior dynamics of neutron stars, forming quantized vortices under rotation. These vortices facilitate angular momentum transfer, driving glitch behavior, while the superfluid density provides a qualitative picture of quantized vortices during the spin-down process. The density profiles shown in Fig.~\ref{fig:denprofile_nr} indicate that the centrifugal potential initially dominates over the circular box trap due to the high rotation frequency. %
As the condensate spins down, the influence of the box trap becomes dominant, resulting in a more uniform density distribution. The presence of the crust significantly increases the vortex number by providing pinning sites within the condensate and also accelerates vortex decay through a depinning mechanism. Liu and colleagues \cite{Liu2025, Liu2024} attribute this behavior to the Magnus flow generated around the pinning sites of the crust.

The Magnus flow, induced by the spin-down of the condensate, reaches a critical velocity that triggers vortex depinning and simultaneous density fluctuations. As the condensate approaches the spin-down time $t_s$, the flow enters a turbulent regime driven by this critical velocity. After spin-down, turbulent fluctuations persist briefly before decaying due to the absence of external forcing.

Further, the vortex avalanche triggered by Magnus flow results in an instantaneous recoupling of the superfluid and normal components, accompanied by a rapid transfer of angular momentum. However, for system with large number of vortices the expulsion happens slowly~\cite{supp}. This abrupt transition in superfluid systems may induce turbulence but cannot be confirmed using density profiles alone. To characterize the turbulent flow in the system, we therefore analyze the incompressible kinetic energy spectra of the condensate. %
\begin{figure}[!ht]
\centering
\includegraphics[width=0.95\linewidth]{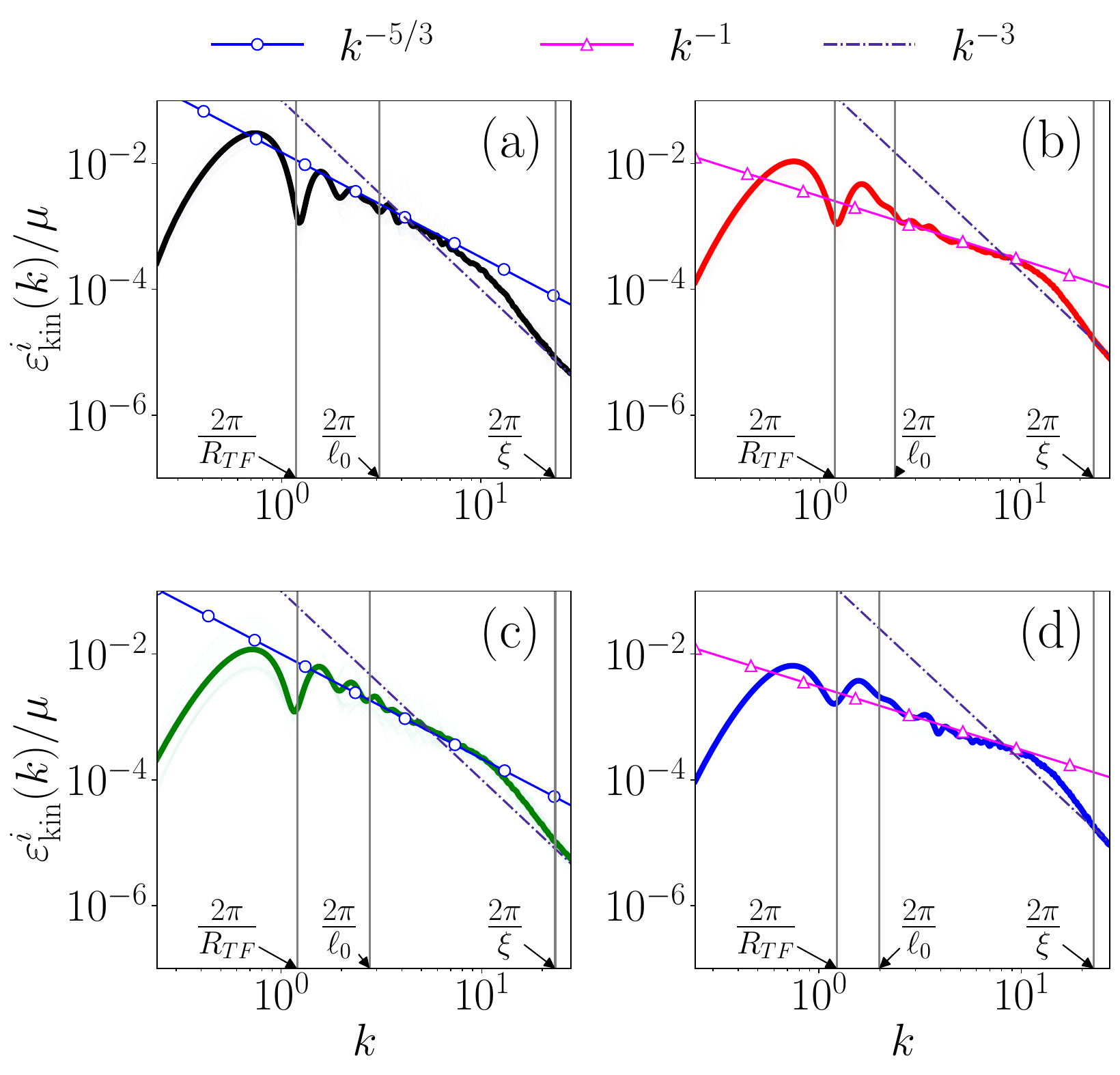}
\caption{Incompressible kinetic energy spectra exhibiting the Kolmogorov cascade for spin-down time ~$t_s = 10$ (a), averaged over $t = 10$ to $t = 30$; (b) averaged over $t = 30$ to $t = 50$, and spin-down time ~$t_s = 20$ (c) averaged over $t = 20$ to $t = 50$; and (d) averaged over $t = 50$ to $t = 60$. The spectra initially [(a) and (c)] and later durations [(b) and (d)] exhibit  $k^{-5/3}$ and $k^{-1}$ scalings, respectively alongside a $k^{-3}$ scaling. }
\label{fig:kolmogorov_vinen}
\end{figure}
We characterize the turbulent regimes by using Thomas-Fermi radius $R_{TF}$, the inter-vortex distance $\ell_0$, and the healing length $\xi$. For two-dimensional systems, the inter-vortex distance is approximated as $\ell_0 = 1/\sqrt{l_v}$, where $l_v$ represents the vortex density per unit area.

In the turbulent regime, the incompressible kinetic energy spectrum exhibits a $k^{-5/3}$ scaling in the inertial range $2\pi/R_{TF} < k < 2\pi/\ell_0$ [see Figs.~\ref{fig:kolmogorov_vinen}(a) and \ref{fig:kolmogorov_vinen}(c)], associated with vortex breakdown. Additionally, the condensate enstrophy undergoes a self-similar cascade, manifested as a $k^{-3}$ power-law scaling at large $k$ values in the incompressible spectrum. For longer spin-down times, the $k^{-5/3}$ scaling weakens, while the $k^{-3}$ scaling becomes more pronounced. The observation that turbulent instability is triggered by rapid transitions in angular frequency and that the energy cascade extends beyond the intervortex spacing aligns with findings on rotational two-dimensional quantum turbulence under varying rotation frequencies \cite{Sivakumar2024}.

Following the initial turbulent behavior exhibiting a Kolmogorov cascade, the vortex system transitions to a Vinen turbulence phase as several vortices decay due to the spin-down. This Vinen turbulence is characterized by a $k^{-1}$ power-law scaling in the range $k > 2\pi/\ell_0$ [see Fig.~\ref{fig:kolmogorov_vinen}(b) and \ref{fig:kolmogorov_vinen}(d)], where individual vortices dominate the velocity field. Notably, the intervortex spacing decreases as vortex density declines, enabling the $k^{-1}$ scaling to manifest at these length scales. Similar to the Kolmogorov regime, the $k^{-1}$ scaling is more pronounced for shorter spin-downs (smaller $t_s$ values). 

As the condensate spins down, the vortex population decreases, reducing the vortex-line density $l_v$. For 2D systems, $l_v$ can be approximated by computing the vortex population over a given area, which leads to increased intervortex spacing $\ell_0$ from the Kolmogorov to the Vinen regime. %
\begin{figure}[!ht]
\centering
\includegraphics[width=0.95\linewidth]{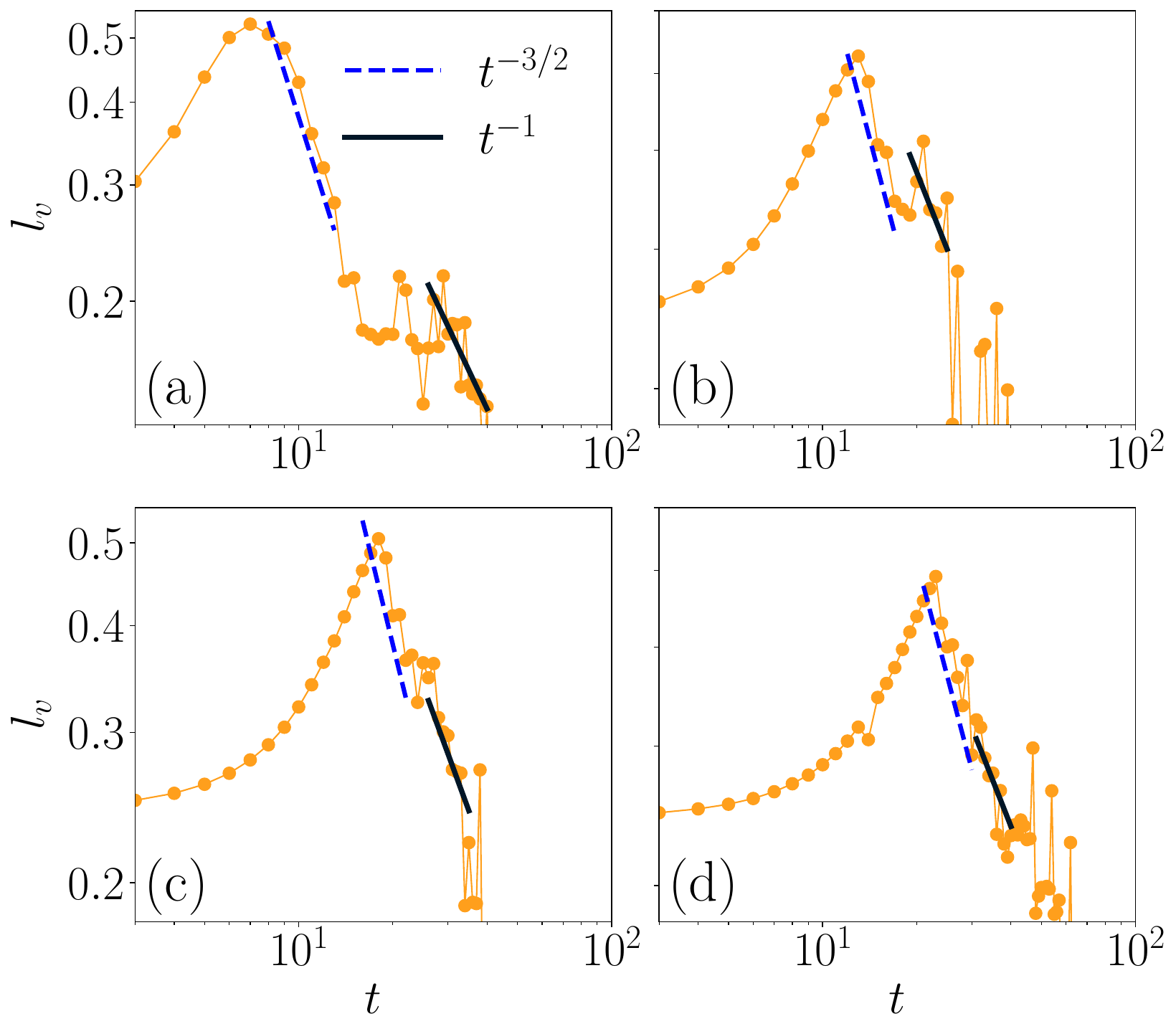}
\caption{Variation of vortex density (per unit area) with respect to time on a logarithmic scale for spin-down times (a)~$t_s = 10$, (b)~$t_s = 20$, (c)~$t_s = 30$, and (d)~$t_s = 40$. The vortex decay in the Kolmogorov and Vinen turbulence regimes exhibits $t^{-3/2}$ and $t^{-1}$ scaling behaviors, respectively.}
\label{fig:nr_decay}
\end{figure}
As shown in Fig.~\ref{fig:nr_decay}, the vortex density exhibits a distinct scaling behavior with respect to time, further confirming the turbulent behavior of the condensate. During the initial Kolmogorov turbulent regime, the vortex-line density $l_v$ scales as $t^{-3/2}$, transitioning to $t^{-1}$ scaling in the later Vinen turbulent regime. This temporal decay of vortex-line density is attributed to vortex breakdown in the Kolmogorov regime, as well as to sound radiation (Kelvin-wave emission) and vortex reconnection in the Vinen regime \cite{Leadbeatera, Leadbeater, Cidrim2017}. For shorter spin-down times, the $t^{-3/2}$ scaling persists over a longer timescale, corresponding to a broader spatial extent of the $k^{-3/2}$ scaling in the incompressible kinetic energy spectrum (Fig.~\ref{fig:kolmogorov_vinen}). {
Due to the intrinsic scales of the Gross–Pitaevskii framework, the power-law behavior observed here is limited to laboratory-scale systems. However, the functional form of this scaling, the $k^{-5/3}$ spectrum, is consistent with the incompressible turbulence spectrum reported in a distinct astrophysical context~\cite{Mocz2017, Sivakumar2025} and system with large number of vortices~\cite{Warszawski2012, supp}. This similarity in spectral behavior, despite the vast difference in physical scales and systems, suggests a common underlying turbulent mechanism may be at play.}

Despite the spatial and temporal profiles of energy and vortex density indicating turbulent behavior, the onset of turbulence after the condensate spins down requires further investigation. We analyze the temporal behavior of the kinetic energy components and observe that the incompressible kinetic energy (associated with vortex flow) initially dominates but decreases as the condensate spins down without rotational forcing to inject vortices into the system.
The quantum pressure energy, which is not associated with the velocity flow of the condensate, becomes dominant over the incompressible component. The spin-down time marks the crossover point where the quantum pressure component surpasses the incompressible component, beyond which Kolmogorov turbulence is observed in the condensate. This finding contrasts with turbulent decay in self-gravitating condensates, as reported in Ref.~\cite{Sivakumar2025}, where the dominance of the quantum pressure component indicates the absence of turbulence. This discrepancy is attributed to differences in the turbulent decay mechanisms. In the self-gravitating case, the decay of incompressible kinetic energy results from the instantaneous expulsion of vortex structures, whereas in the spin-down case, it arises from more gradual vortex decay. Despite reaching the spin-down time, vortices persist in the condensate (see Fig.~\ref{fig:denprofile_nr}) for a sufficient duration to undergo an energy cascade and exhibit turbulent behavior. %
\begin{figure}[!ht]
\centering
\includegraphics[width=\linewidth]{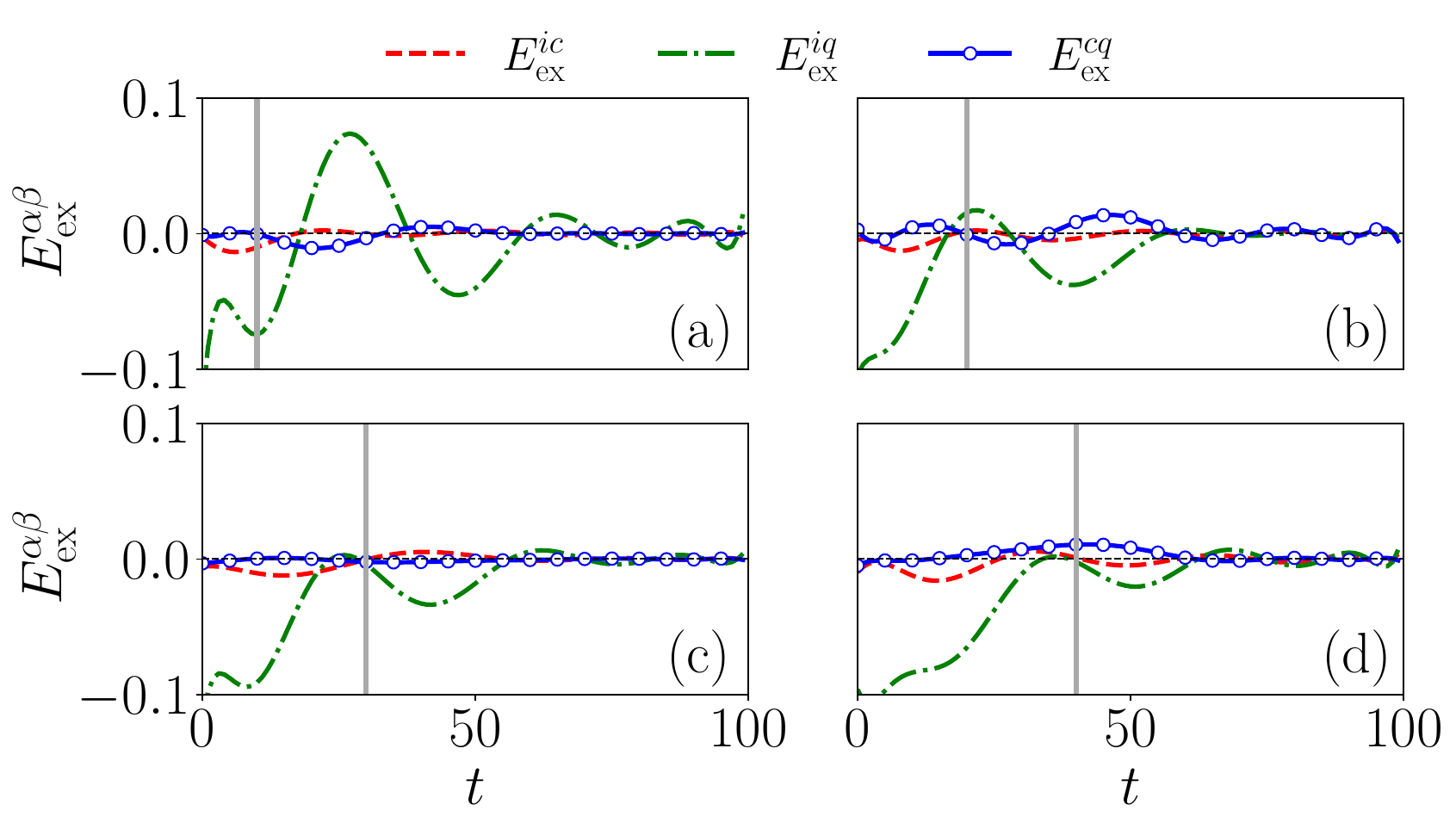}
\caption{Temporal profiles of the total kinetic energy exchange between its components for spin-down times(marked by dark-grey vertical lines) (a) $t_s = 10$, (b) $t_s = 20$, (c) $t_s = 30$, and (d) $t_s = 40$. The red dashed line represents energy transfer from the incompressible to the compressible component ($E_{\mathrm{ex}}^{ic}$), the green dash-dotted line represents energy transfer between the incompressible and quantum pressure components ($E_{\mathrm{ex}}^{iq}$), and the blue line with circles represents energy transfer between the compressible and quantum pressure components ($E_{\mathrm{ex}}^{cq}$). A negative exchange energy indicates a transfer from the first component $\alpha\in\{i,c,q\}$ in the superscript to the second $\beta \in\{i,c,q\}$, while a positive value indicates the reverse.}
\label{fig:nr_exchange}
\end{figure}

Although glitch events are explained by the transfer of angular momentum \cite{ANDERSON1975}, analyzing the onset of turbulence and the dynamics of quantized vortices is required to fully understand the transport of kinetic energy components, which, in turn, enables a new perspective on vortex depinning.
The evolution of net energy transport between the kinetic components, shown in Fig.~\ref{fig:nr_exchange}, provides critical insight into the post-spin-down dynamics and explains the persistence of vortex flow. The energy exchange between the compressible--quantum pressure and incompressible--compressible components does not significantly influence the turbulent dynamics. However, the energy exchange between the incompressible and quantum pressure components, denoted $E_{\mathrm{ex}}^{iq}$, is initially negative, indicating energy transfer from the incompressible to the quantum pressure component. After the condensate spins down, the exchange becomes positive, indicating a transfer from quantum pressure to incompressible component and at later times $E_{\mathrm{ex}}^{iq} \to 0$, as no further incompressible energy is transferred to quantum pressure. For sufficiently short spin-down times [see Fig.~\ref{fig:nr_exchange}(a)--(b)], the exchange energy becomes positive, indicating an injection of incompressible energy from the quantum pressure component. This secondary injection mechanism sustains turbulent behavior in the condensate, even after the loss of rotational forcing, which explains the existence of vortex flow after the spin-down time is reached. For longer spin-down times, the peak of the secondary injection diminishes, reducing the intensity of turbulence.

The damping coefficient $\gamma$ is used in atomic BECs to model vortex-sound interactions~\cite{Bradley2008}. A comparable approach is employed in neutron star interiors to describe interactions with the outer crust and proton superconducting core. In addition, the impact of $\gamma$ on vortex and turbulent dynamics under the damped GP model provides an important clue to the contribution of the interaction between neutron star interiors, which are mainly composed of superfluid and proton superconductors. For atomic BECs, the damping parameter $\gamma$ can be controlled by varying the temperature of the condensate, significantly affecting the vortex and density-wave dynamics. The damping parameter $\gamma$ is controlled by varying the temperature of the condensate, significantly affecting vortex and density-wave dynamics. %
\begin{figure}[htp]
\centering
\includegraphics[width=\linewidth]{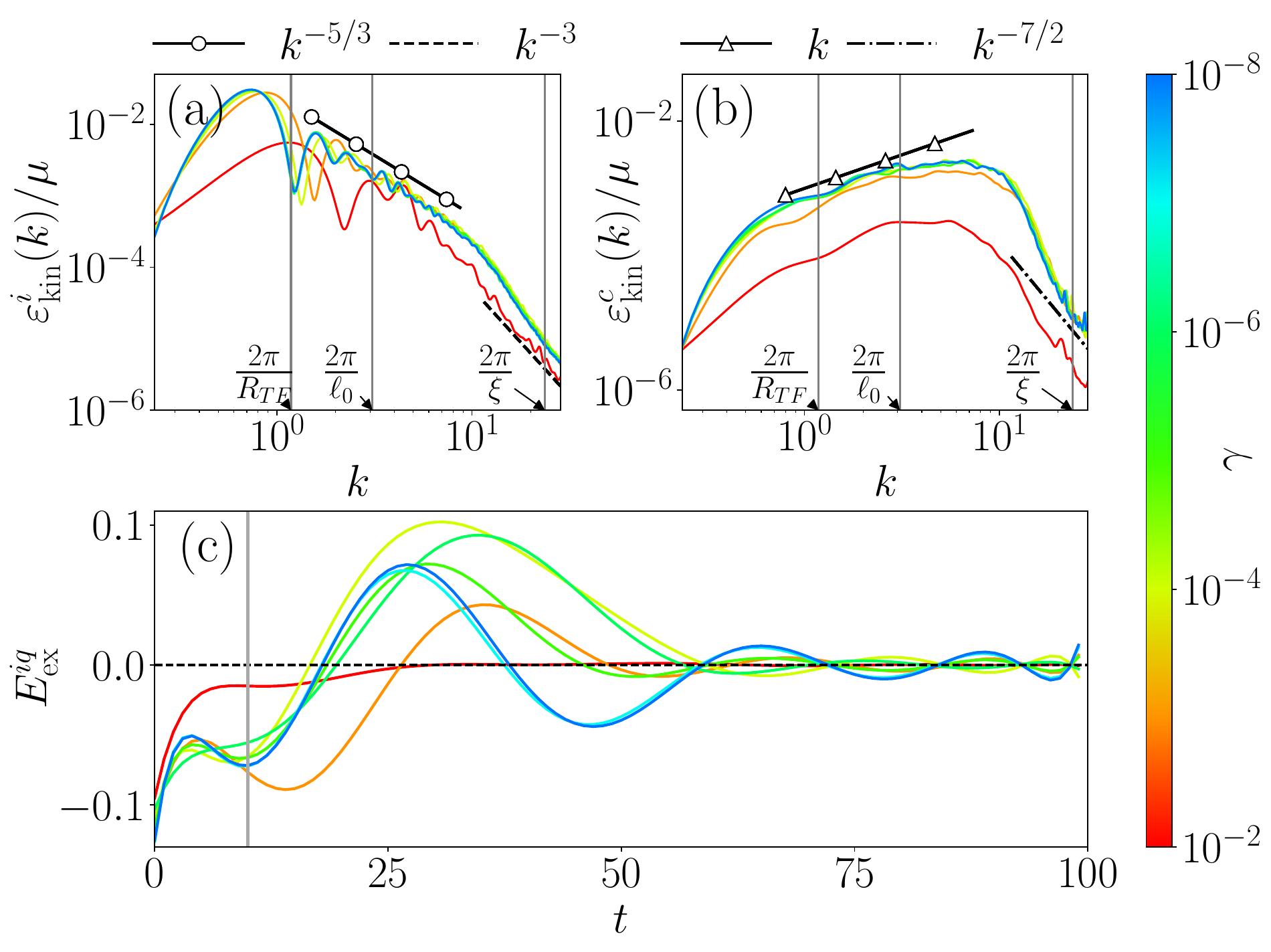}
\caption{Variation of (a) incompressible kinetic energy spectra, (b) compressible kinetic energy spectra, and (c) Time evolution of the exchange energy between the incompressible and quantum pressure components for spin-down time $t_s=10$ (dark-gray vertical line), for different values of the damping coefficient $\gamma$.} 
\label{fig:spec_gamma}
\end{figure}%
Figure~\ref{fig:spec_gamma}(a-b) illustrates the evolution of incompressible and compressible kinetic energy spectra for various values of the damping coefficient $\gamma$. For the strongly turbulent case with $t_s = 10$ [see Figs.~\ref{fig:kolmogorov_vinen}(a)], the spectra exhibit pronounced $k^{-5/3}$ scaling in the incompressible component and $k$ scaling in the compressible component for smaller damping values (underdamped regime). The $k$ scaling in the infrared region and $k^{-7/2}$ scaling in the ultraviolet regime for the compressible spectra indicates the presence of weak wave turbulence~\cite{Estrada2022, Sivakumar2024a, Sivakumar2025}. At approximately $\gamma \approx 10^{-2}$, the spectra abruptly lose their scaling behavior as stronger damping (overdamped regime) dissipates the turbulent flow. This turbulent flow is further confirmed by the emergence of a prominent $k^{-3}$ scaling and the diminishing $k^{-5/3}$ scaling in the overdamped case ($\gamma \approx 10^{-2}$), indicating the inhibition of vortex breakdown and the formation of a stable vortex lattice structure. It is worthwhile to mention mainly an over-damped condensate ($\gamma > 0.01$) has been considered to model superfluid cores of neutron stars due to several potential dissipation pathways~\cite{Warszawski2012, Loennborn2019}.

Similar to the non-damped case, we complement the spectral analysis with the exchange energy profile in the presence of damping Fig.~\ref{fig:spec_gamma}(c). The injection of incompressible kinetic energy reaches a maximum at an optimal value of $\gamma$, situated between the underdamped and overdamped regimes. Under these optimal damping conditions, the transfer of energy from quantum pressure to incompressible modes appears to be significantly enhanced. Given that $\gamma$ serves as a simplified parametrization of dissipation in quantum fluids, the underlying mechanism responsible for the observed maximum in incompressible energy injection at this intermediate damping requires further investigation. In the overdamped regime, around $\gamma \approx 10^{-2}$, the positive peak in exchange energy is absent, consistent with the suppression of turbulence under such strongly dissipative conditions.

\textit{Summary and Conclusion: }
We investigated two-dimensional quantum turbulence in a spinning-down condensate subjected to a crust potential that provides vortex pinning. Continuous spin-down profiles show that as the spin-down duration $t_s$ approaches the quasi-discrete limit, turbulent behavior emerges. After spin-down, the system exhibits an incompressible kinetic energy spectrum with Kolmogorov scaling $k^{-5/3}$ and a compressible spectrum scaling as $k^{1}$, indicating thermalization. The flow subsequently enters the Vinen regime, dominated by isolated vortices and characterized by a $k^{-1}$ scaling. In both regimes, spectral scaling weakens as $t_s$ increases, while the vortex-line density $l_v$ decays as $t^{-3/2}$ (Kolmogorov) and $t^{-1}$ (Vinen).

The interplay between kinetic energy components reveals that spin-down drives a transfer from incompressible kinetic energy to quantum pressure. After spin-down, a reverse transfer from quantum pressure back to incompressible energy triggers a turbulent cascade even in the absence of external forcing. This secondary energy injection appears as an increase in vortex-line density near $t_s$ and is attributed to vortex depinning and avalanche events in the neutron star cores~\cite{Khomenko2018, Loennborn2019, Howitt2022}.

A sharp transition from non-turbulent to turbulent scaling occurs near a damping parameter $\gamma \sim 10^{-2}$, with an optimal value maximizing energy transfer from quantum pressure to incompressible modes. For longer spin-down durations, vortex avalanches have been associated with pulsar glitches \cite{Liu2025}. Since quantum turbulence suppresses collective excitations \cite{Lee2025, Ferrand2021}, the turbulence onset conditions identified here may be relevant to glitch formation in pinned systems. In addition to depinning-driven glitches, self-gravitating BECs without pinning can expel vortices once the circulation exceeds a critical threshold $\xi_c$, potentially producing glitch-like behavior via crust interaction \cite{Nikolaieva2021, Sivakumar2025}.

These findings complement the already existing Gross-Pitaevskii models of neutron star cores~\cite{Shukla, Verma, Liu2025}. Mapping turbulence dynamics across spin-down profiles and dissipation parameters highlights both the applicability and limitations of the GP analogy for neutron star glitch phenomenology \cite{Liu2025}. The results also advance understanding of inhomogeneous two-dimensional quantum turbulence, highlighting secondary energy injection driven by repeated vortex pinning and depinning.

Although the GP framework simplifies neutron superfluid microphysics, it remains the most microscopically faithful model for quantized vortex dynamics. While extrapolation to neutron star scales is not yet quantitative, the robust appearance of vortex avalanches, power-law glitch statistics, and rapid rise times across GP studies supports the astrophysical relevance of the underlying mechanisms. The transient quantum-pressure-driven energy-injection process identified here thus represents a potentially important ingredient for future multi-scale models of neutron star interiors.

\appendix
\section{Energy Injection in Larger Condensates with More Vortices}

\label{sec:appendix_a}
In this appendix, we provide additional details on the energy injection mechanism in larger condensates supporting a greater number of vortices. To further investigate this mechanism in larger condensates containing a greater number of vortices (e.g., approximately \(70\)), we performed simulations involving a partial spin-down of the condensate, reducing the frequency from $\Omega = 2.0$ to $\Omega = 1.0$. Under these conditions, we observed that vortex expulsion does not occur simultaneously; instead, vortices are shed progressively.
\begin{figure}[!ht]
    \centering
    \includegraphics[width=0.99\linewidth]{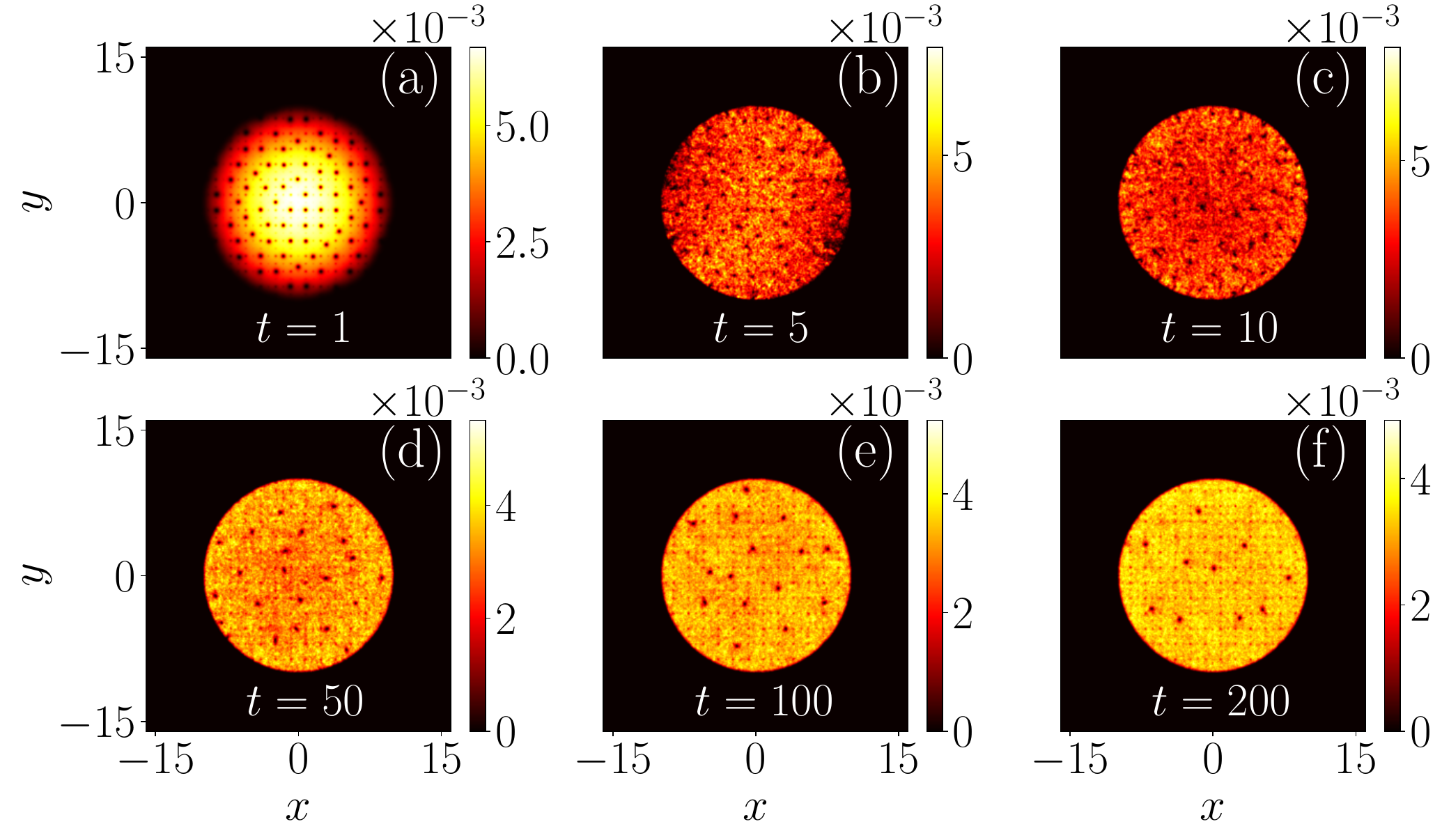}
    \caption{Snapshots of condensate density at various times for a larger condensate rotating from $\Omega=2.0$ to $\Omega=1.0$ with spin-down time $t_s=10$.}
    \label{fig:author:density}
\end{figure}
Figure~\ref{fig:author:density} presents snapshots of the condensate density at various time instances during a spin-down of duration \(10\) units. In this process, vortices near the condensate center must travel longer distances to reach the edge, increasing the likelihood of multiple pinning–depinning events. This behavior significantly inhibits the vortex avalanche mechanism, thereby reducing the size of glitches in condensates with larger vortex populations. 

\begin{figure}[ht!]
    \centering
    \includegraphics[width=0.99\linewidth]{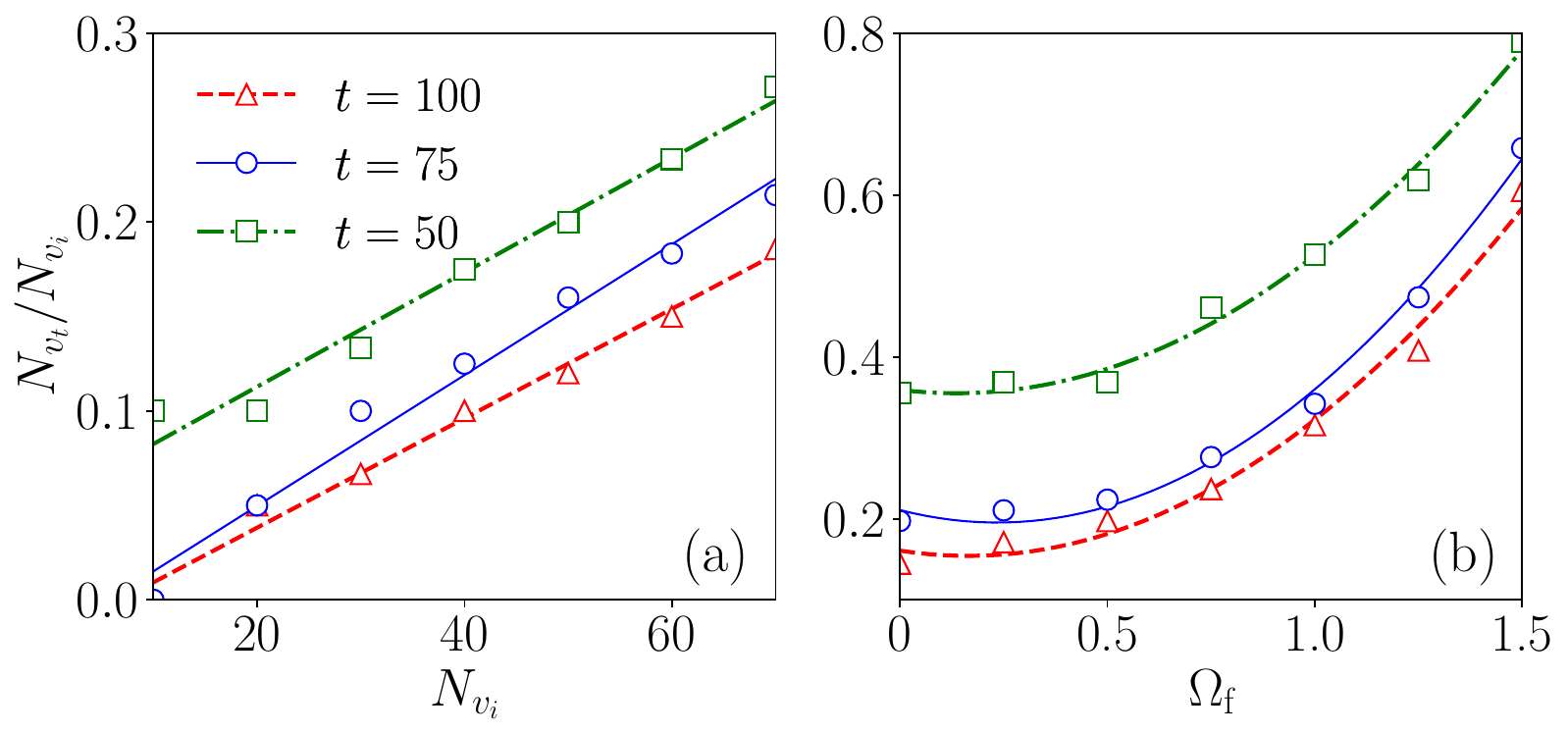}
    \caption{Vortex fraction of the condensate after times $t = 100, 75$ and $t=50$, with respect to (a) initial vortex population for a condensate and (b) final rotation frequency $\Omega_{\mathrm{f}}$. In both cases the vortex fraction is higher at later duration, indicating a inhibition of vortex expulsion.}
    \label{fig:vortex_fraction}
\end{figure}

This inhibition is further corroborated by the ratio of vortices at time $t$ ($N_{v_t}$)to the initial vortex population ($N_{v_i}$), whose evolution with respect to the initial vortex number is illustrated in Fig.~\ref{fig:vortex_fraction}(a). Since Bose-Einstein condensates (BECs) maintain a uniform areal density, the vortex population serves as a reliable proxy for condensate size. Figure~\ref{fig:vortex_fraction}(a) shows a linear increase in vortex retention with increasing condensate size. This observation provides a physical explanation for the upper bound on the number of vortices required to observe glitch-like behavior, as reported in Ref.~\cite{Warszawski2011}.

Additionally, the figure indicates that the vortex fraction at $t=75$ and $t=100$ is nearly identical, suggesting that the vortex expulsion process significantly slows at later times. In Fig.~\ref{fig:vortex_fraction}(b), the vortex fraction for a larger condensate exhibits an exponential increase with respect to the final rotation frequency $\Omega_{\mathrm{f}}$. For smaller $\Omega_{\mathrm{f}}$, the vortex fraction remains nearly constant, indicating that vortex expulsion is comparably effective for both complete spin-down  ($\Omega_{\mathrm{f}}$ = 0) and partial spin-down to lower rotation frequencies ($\Omega_{\mathrm{f}}< 0.5$ ).

\begin{figure}[!ht]
    \centering
    \includegraphics[width=0.99\linewidth]{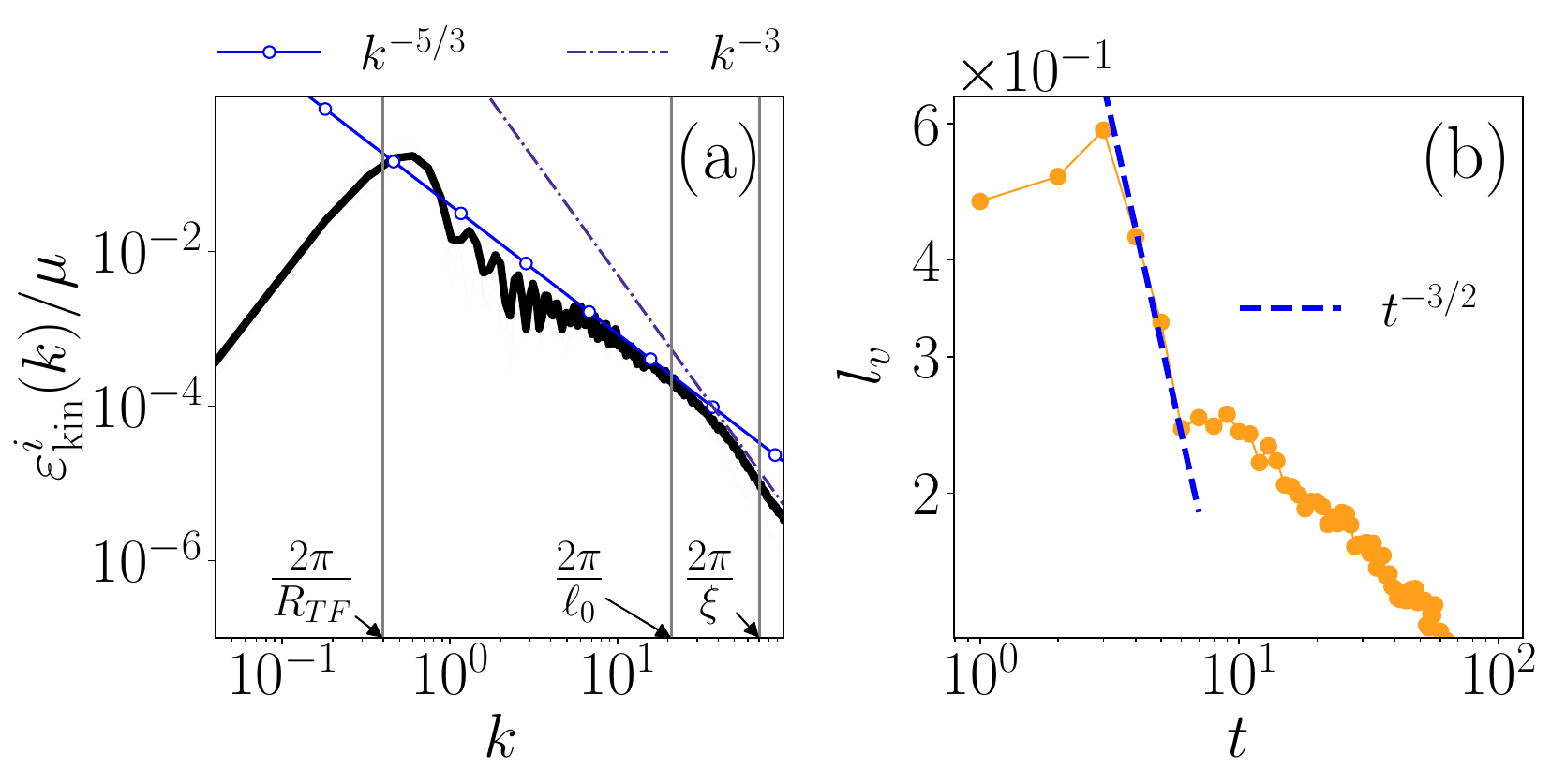}
    \caption{(a) Incompressible kinetic energy spectra displaying $k^{-5/3}$ and $k^{-3}$ scaling averaged over time $t=5$ to $t=10$, indicating Kolmogorov cascade in the larger condensate with partial spin-down. (b)Vortex density (per unit area) as a function of time displaying a scaling behavior as $t^{-3/2}$ consistent with the  Kolmogorov-like cascade.}
    \label{fig:author:spectra}
\end{figure}

Despite the inhibited avalanche activity, the larger vortex system still exhibits turbulence before reaching the final rotation frequency. Figure~\ref{fig:author:spectra}(a) shows the incompressible kinetic energy spectra, averaged over the interval \(t = 5\) to \(t = 10\), displaying \(k^{-5/3}\) and \(k^{-3}\) scaling, characteristic of the Kolmogorov cascade. %
Figure~\ref{fig:author:spectra}(b) shows the time evolution of vortex density on a logarithmic scale. Similar to the smaller condensate case, a $t^{-3/2}$ scaling is observed during the turbulent regime. However, after this phase, vortex density decays more slowly than in smaller systems, indicating a prolonged persistence of vortices following spin-down. This slow decay further supports the conclusion that inter-vortex spacing and vortex core size do not directly influence the turbulent dynamics in our Gross–Pitaevskii model, as the vortex areal density remains of the same order of magnitude across different system sizes.

It is important to note that turbulence arises before the completion of the spin-down. %
\begin{figure}[!ht]
    \centering
    \includegraphics[width=0.99\linewidth]{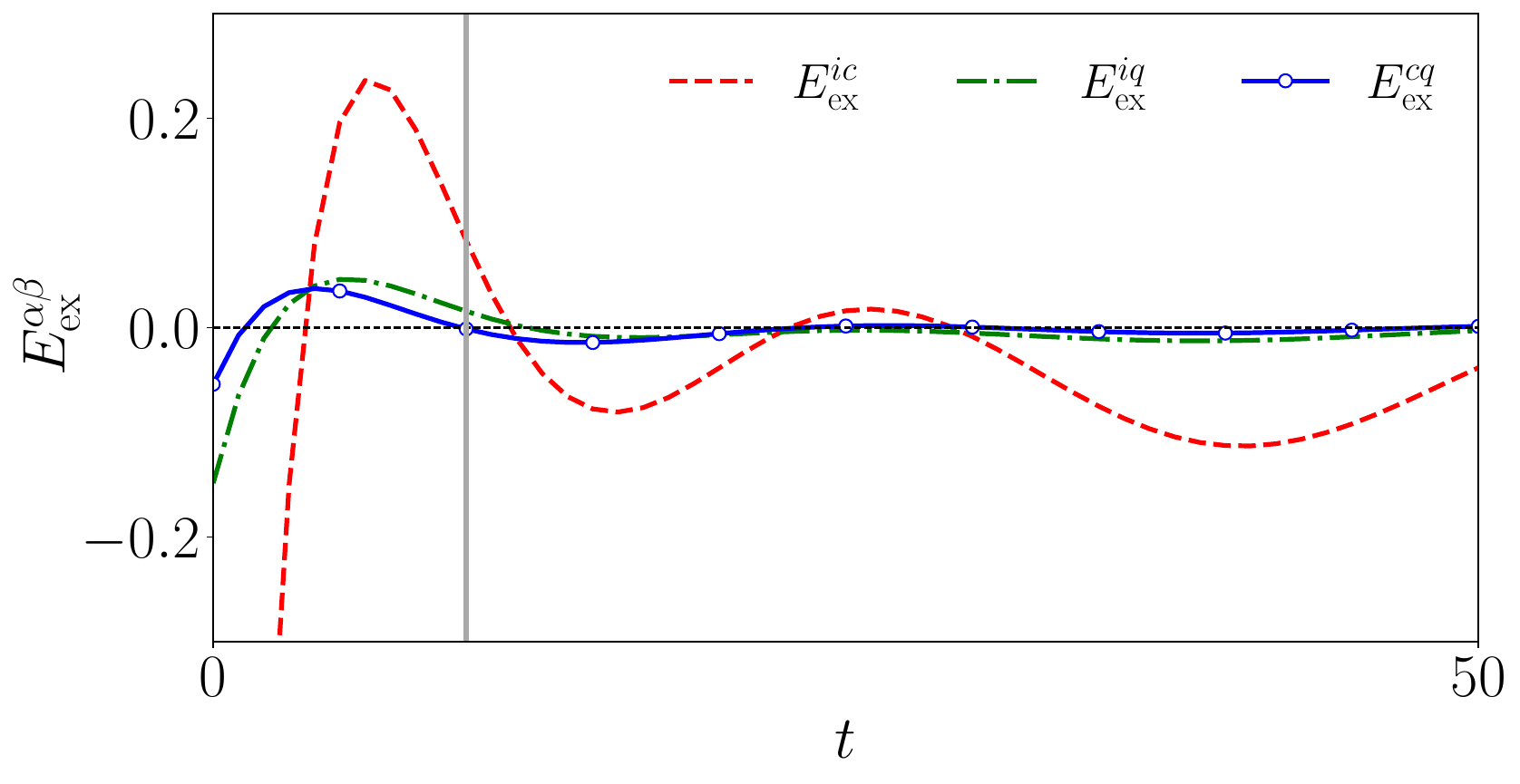}
    \caption{Evolution of energy exchange between compressible-incompressible components (red dashed line), quantum pressure-incompressible components (green dash-dotted line), and quantum pressure-compressible components (blue circled line). For larger condensates, transitioning between two non-zero frequencies, compressible to incompressible exchange dominates over the quantum pressure to incompressible exchange due to the persistence of vortices.}
    \label{fig:author:exchange}
\end{figure}
This is attributed to the larger number of vortices and the enhanced compressible flow in the system, which together increase the likelihood of de-pinning events occurring before the target rotation frequency is reached---unlike in the smaller condensate case.

To further analyze the energy dynamics during this process, we plot the evolution of energy exchange between the compressible and incompressible components, the quantum pressure and incompressible components, and the quantum pressure and compressible components in Fig.~\ref{fig:author:exchange}. For larger condensates undergoing a transition between two non-zero rotation frequencies, energy transfer from the compressible to the incompressible component dominates over the quantum pressure–to–incompressible transfer. Notably, this energy injection from the compressible flow coincides with the onset of the Kolmogorov cascade, indicating its key role in driving turbulence. The suppression of quantum pressure–to–incompressible transfer in this regime is primarily due to the persistence of vortices and the enhanced generation of sound waves, both of which are facilitated by the larger condensate size.

In contrast, smaller condensates generate fewer sound waves, and vortex expulsion is more pronounced during the spin-down process. As a result, the compressible–to–incompressible transfer is weaker or absent during the turbulent phase, making the quantum pressure the dominant source of incompressible energy injection. Furthermore, vortex expulsion in larger condensates is significantly suppressed, even when the condensate is fully spun down, due to the system’s extended spatial scale. Vortices persist for long durations and are only expelled gradually over time. Since Bose-Einstein condensates sustain vortices only above a certain threshold rotation frequency, a rapidly rotating condensate can remain vortex-free if the interaction strength and trapping potential are tuned appropriately. In such cases, the system can effectively behave like a non-rotating condensate.

Our simulations provide clear evidence of the simultaneous onset of turbulence and energy injection before the system reaches the lower rotation frequency. Most importantly, incompressible energy input from quantum pressure is masked by the more dominant compressible–to–incompressible transfer. These effects stem from the increased system size and partial spin-down, both of which hinder vortex expulsion and prolong the turbulent dynamics.

In the present model, certain aspects, such as crust potentials and related parameters, are treated phenomenologically, reflecting the diversity of existing models. Although a universal scaling law for glitch behavior across condensate sizes remains elusive, our simulations reveal qualitative consistency with turbulent scaling in larger condensates subjected to partial spin-down. Under these conditions, however, the contribution of quantum pressure to incompressible energy injection is notably reduced. This observation suggests that increased vortex expulsion enhances quantum pressure-driven energy transfer, highlighting the interplay between vortex dynamics and energy injection mechanisms.

\acknowledgments

A.S. acknowledges financial support from the Council of Scientific and Industrial Research (CSIR), India, in the form of a Direct Senior Research Fellowship. The work of P.M. is supported by the Ministry of Education-Rashtriya Uchchatar Shiksha Abhiyan (MoE RUSA 2.0): Bharathidasan University -- Physical Sciences. 


\begin{thebibliography}{50}
\bibitem [{sup()}]{supp}%
  \BibitemOpen
  \href@noop {} {}\bibinfo {note} {See Appendix \ref{sec:appendix_a} for details on the energy injection mechanism via partial spin-down in a larger condensate that supports a greater number of vortices.}\BibitemShut {Stop}%
\end{thebibliography}

%

\end{document}